\documentclass[prd,aps,twocolumn,a4paper,floatfix,nofootinbib,showpacs]{revtex4}
   
\usepackage{graphicx,psfrag}
\usepackage{mathrsfs}
\usepackage{amsmath,amsfonts,amssymb}
\usepackage{multirow}
\usepackage{comment}
\usepackage[colorlinks=true]{hyperref}
\hypersetup{
  linkcolor=blue,
  citecolor=blue,
  urlcolor=blue
}
\usepackage{bbm}
\usepackage{placeins}
\usepackage[normalem]{ulem}
\usepackage{tensor} 
\usepackage{acronym}
\usepackage{xspace}
\usepackage{booktabs,multirow,cellspace}

\usepackage{color}
\definecolor{cyan}{rgb}{0,0.9,0.9}
\definecolor{orange}{rgb}{0.9,0.5,0}
\definecolor{purple}{rgb}{0.8,0.4,0.8}
\definecolor{gray}{rgb}{0.8242,0.8242,0.8242}
\definecolor{pink}{rgb}{1.0, 0.0, 0.5}
\newacro{BBH}{binary-black-hole}
\newacro{BH}{black hole}
\newacro{BNS}{binary neutron star}
\newacro{EOB}{effective-one-body}
\newacro{EOS}{equation-of-state}
\newacro{GW}{gravitational-wave}
\newacro{NR}{numerical-relativity}
\newacro{NS}{neutron star}
\newacro{PN}{post-Newtonian}
\newacro{PSD}{power spectral density}
\newacroplural{PSD}[PSDs]{power spectral densities}
\newacro{SNR}{signal-to-noise ratio}
\newacro{ODE}{ordinary-differential-equation}
\newacro{ZDHP}{zero-detuned high-power configuration}
\def\NRtidal{\texttt{NRTidal}\xspace}
\def\SEOBNRROM{\texttt{SEOBNRv4\_ROM}\xspace}
\def\SEOBNRROMNRtidal{\texttt{SEOBNRv4\_ROM\_NRTidal}\xspace}
\def\IMRPhenomDNRtidal{\texttt{IMRPhenomD\_NRTidal}\xspace}
\def\IMRPhenomD{\texttt{IMRPhenomD}\xspace}
\def\TEOBResum{\texttt{TEOBResum}\xspace}
\def\TEOBResumROM{\texttt{TEOBResum\_ROM}\xspace}

\def\TaylorF{\texttt{TaylorF2}\xspace}
\def\TaylorFT{\texttt{TaylorF2$_{\rm Tides}$}\xspace}

\definecolor{mygreen}{rgb}{0.1,0.8,0.1}

\begin{document}

\title{Relevance of tidal effects and post-merger dynamics for binary neutron star parameter estimation}

\author{Reetika \surname{Dudi$^1$}, Francesco \surname{Pannarale$^{2,3}$}, Tim \surname{Dietrich$^{4}$}, Mark \surname{Hannam$^2$}, 
Sebastiano \surname{Bernuzzi$^{1,5}$}, Frank \surname{Ohme$^{6,7}$}, Bernd \surname{Br\"ugmann$^1$}}

\affiliation{$^1$Theoretical Physics Institute, University of Jena, 07743 Jena, Germany}
\affiliation{$^2$Gravity Exploration Institute, School of Physics and Astronomy, Cardiff University, The Parade, Cardiff CF24 3AA, UK}
\affiliation{$^3$Dipartimento di Fisica, Universit\`a di Roma ``Sapienza'' \& Sezione INFN Roma1, P.A.\,Moro 5, 00185, Roma, Italy}
\affiliation{$^4$ Nikhef, Science Park, 1098XG Amsterdam, The Netherlands}
\affiliation{$^5$Istituto Nazionale di Fisica Nucleare, Sezione Milano Bicocca, gruppo collegato di Parma,
Parco Area delle Scienze 7/A, I-43124 Parma, Italy}
\affiliation{$^6$Max Planck Institute for Gravitational Physics (Albert Einstein Institute), Callinstra{\ss}e 38,
30167 Hannover, Germany}
\affiliation{$^7$Leibniz Universit\"at Hannover, 30167 Hannover, Germany}

\date{\today}

\begin{abstract} 
  Measurements of the properties of binary neutron star systems from
  gravitational-wave observations require accurate theoretical models
  for such signals. However, current models are incomplete, as they do
  not take into account all of the physics of these systems: some
  neglect possible tidal effects, others neglect spin-induced orbital
  precession, and no existing model includes the post-merger regime
  consistently. In this work, we explore the importance of two
  physical ingredients: tidal interactions during the inspiral and the
  imprint of the post-merger stage. We use complete
  inspiral--merger--post-merger waveforms constructed from a tidal
  effective-one-body approach and numerical-relativity simulations as
  signals against which we perform parameter estimates with waveform
  models of standard LIGO-Virgo analyses. We show that neglecting
  tidal effects does not lead to appreciable measurement biases in
  masses and spin for typical observations (small tidal deformability
  and signal-to-noise ratio $\sim$ 25). However, with increasing
  signal-to-noise ratio or tidal deformability there are biases in the
  estimates of the binary parameters. The post-merger regime, instead,
  has no impact on gravitational-wave measurements with current
  detectors for the signal-to-noise ratios we consider.
\end{abstract}


\maketitle

\section{Introduction}\label{Section:Introduction}
On August 17, 2017, the LIGO-Virgo
collaboration~\cite{TheLIGOScientific:2014jea, TheVirgo:2014hva}
observed for the first time a \ac{GW} signal consistent with a
\ac{BNS} coalescence~\cite{TheLIGOScientific:2017qsa}. The signal,
GW170817, was detected with a combined \ac{SNR} of $32.4$, making it
the strongest \ac{GW} signal observed to date, with the source located
at a luminosity distance of only $40^{+8}_{-14}$\,Mpc from Earth.
This observation was associated with the short gamma-ray burst event
GRB 170817A, confirming \ac{BNS} mergers as a progenitor for short
gamma-ray bursts~\cite{monitor:2017mdv}. Further, it sparked a global
electromagnetic follow-up campaign (see~\cite{GBM:2017lvd} and
references therein) and led to an independent measurement of the
Hubble constant~\cite{Abbott:2017xzu}, as well as new constraints on
the \ac{NS} \ac{EOS}~\cite{Annala:2017llu,Fattoyev:2017jql,
  Nandi:2017rhy, De:2018uhw, Abbott:2018exr}. These results and the
extraction of binary properties in general~\cite{Abbott:2018wiz},
including the tidal deformability of the stars, rely on Bayesian
inference methods that compare the observed signal against theoretical
models~\cite{Veitch:2014wba, Biwer:2018pyc}. A complete analysis of
the source parameters estimated from GW170817 is given in
Ref.~\cite{Abbott:2018wiz}.

The fidelity of parameter measurements depends on detector calibration
uncertainty \cite{Viets:2017yvy, Cahillane:2017vkb, Acernese:2018bfl},
detector performance at the time of the event (both in terms of the
overall sensitivity to the signal, and of the stability of the
instrument due to the presence/absence of transient noise fluctuations
\cite{TheLIGOScientific:2016zmo}), systematic errors in the
theoretical waveforms employed to analyze the data, and any signal
correlations between source parameters.  Here we focus on systematic
errors due to approximations or missing physics in the waveform
models.  As opposed to the case of the first \ac{BBH}
observation~\cite{Abbott:2016blz, TheLIGOScientific:2016wfe,
  Abbott:2016izl, abbott:2016wiq}, where full \ac{BBH}
inspiral-merger-ringdown waveform models were used for parameter
estimation, GW170817 was analysed using models with different
approximate treatments of tidal effects, and no model described the
system post-merger~\cite{Abbott:2018wiz}.

Several studies investigated the measurability of the \ac{NS} tidal
deformability or the detectability of the post-merger signal in the
case of \ac{BNS} coalescence observations, but a full Bayesian
analysis with complete waveforms has not been performed to date.
Flanagan and Hinderer~\cite{Flanagan:2007ix} considered the early (up
to 400\,Hz) inspiral of \ac{PN} waveforms and showed that advanced
detectors could constrain the \ac{NS} tidal deformability for a
putative source at $50$\,Mpc. Hinderer et al.~\cite{Hinderer:2009ca}
investigated the possibility of using such constraints on the tidal
deformability to distinguish among \ac{NS} \ac{EOS} models.  They
found that advanced detectors would probe only unusually stiff
\acp{EOS}, while the Einstein Telescope could provide a clean \ac{EOS}
signature.  Damour et al.~\cite{Damour:2012yf} studied tidally
corrected \ac{EOB} waveforms up to merger and concluded that an
advanced detector network could measure \ac{NS} tidal polarizability
parameters from \ac{GW} signals at an \ac{SNR} of
$16$. Favata~\cite{Favata:2013rwa} investigated the accuracy with
which masses, spins, and tidal Love numbers can be constrained in the
presence of systematic errors in waveform models.  He found that
neglecting spins, eccentricity, or high-order \ac{PN} terms could
significantly bias measurements of \ac{NS} tidal Love numbers.

All studies summarized above relied on the Fisher matrix
approximation, which holds for loud signals.  Making strong statements
about estimating source parameters requires a full Bayesian analysis.
This was carried out for the first time with tidally corrected \ac{PN}
waveforms by Del Pozzo et al.~\cite{DelPozzo:2013ala}. They showed
that second generation detectors could place strong constraints on the
\ac{NS} \ac{EOS} by combining the information from tens of detections.
A full Bayesian analysis in the case of advanced detectors was also
carried out by Wade et al.~\cite{Wade:2014vqa} who found that
systematic errors inherent in the \ac{PN} inspiral waveform families
significantly bias the recovery of tidal parameters. Lackey and
Wade~\cite{Lackey:2014fwa} provided a method to estimate the \ac{EOS}
parameters for piecewise polytropes by stacking tidal deformability
measurements from multiple detections, also concluding that a few
bright sources would allow one to the \ac{NS} \ac{EOS}. Agathos et
al.~\cite{Agathos:2015uaa} revisited the problem of distinguishing
among stiff, moderate and soft \acp{EOS} using multiple detections. In
contrast to~\cite{DelPozzo:2013ala, Wade:2014vqa}, they used a large
number of simulated \ac{BNS} signals and took into account more
physical ingredients, such as spins, the quadrupole-monopole
interaction, and tidal effects to the highest (partially) known order.
Later, \citet{Chatziioannou:2018vzf} extended the work
of~\cite{Wade:2014vqa} using a more appropriate spectral \ac{EOS}
parametrization, while \citet{Carney:2018sdv} showed that imposing a
common \ac{NS} \ac{EOS} leads to improved tidal
inference. \citet{Chatziioannou:2015uea} considered the problem of
using \ac{BNS} inspirals to distinguish among \acp{EOS} with different
internal composition and concluded that the existence/absence of
strange quark stars is the most straightforward scenario to probe with
second generation detectors.  They also showed that stacking multiple
moderately low \ac{SNR} detections should be carried out with caution
as the procedure may fail when the prior information dominates over
new information from the data. Finally, Clark and
collaborators~\cite{Clark:2014wua, Clark:2015zxa} provided the first
systematic studies of the detectability of high-frequency content of
the merger and post-merger part of \ac{BNS} \ac{GW} signals.  As
opposed to the studies outlined previously, these investigations did
not rely on waveform models and optimal filtering, but exploited
methodologies used to search for unmodelled \ac{GW} transients.  They
focused on the problem of discriminating among different post-merger
scenarios and on measuring the dominant oscillation frequency in the
post-merger signal, concluding that second generation detectors could
detect post-merger signals and constrain the \ac{NS} \ac{EOS} for
sources up to a distance of 10--25\,Mpc (assuming optimal
orientation).

In this article, we focus on two sources of systematic uncertainties
and try to answer the following questions.

(i) { \it What is the impact of neglecting tidal effects in the
  analysis of the inspiral \ac{GW} signal}?

As the two \acp{NS} orbit and slowly inspiral, each one becomes
tidally deformed by the gravitational field of its companion.  This
effect leads to an increase in the inspiral
rate~\cite{Flanagan:2007ix}. The inspiral rate also increases if the
angular momenta of the bodies, {\it i.e.}, the spins, are aligned in
the opposite direction to the orbital angular momentum of the binary,
or by a change in the binary mass-ratio~\cite{Baird:2012cu}; it is
plausible, then, that neglecting tidal effects could lead to biases in
mass and spin measurements.  The extent of this effect will depend on
how easily the \acp{NS} can be deformed. In \ac{PN} calculations of
binary inspirals, tidal effects enter at high (5\ac{PN})
order~\cite{Damour:1983a, Damour:1992qi, Racine:2004hs,
  Flanagan:2007ix, Damour:2009wj, Vines:2011ud}, so for weak signals
or small tidal deformabilities, it is possible that tidal effects
could be neglected when measuring source properties such as masses and
spins. We find that this is true for \acp{SNR} at least as high as 25,
for \acp{EOS} consistent with current observations. If the tidal
deformability is larger, or the signal has a much higher \acp{SNR},
then neglecting tidal terms would lead to a bias in other source
parameters. We will show examples of this within the article.

(ii) {\it Does the use of inspiral-only waveforms lead to a
  significant loss of information, or possibly to biases in the
  estimation of the source properties?}

Currently, waveform models used to interpret \ac{BNS} observations do
not include the merger and post-merger regimes. Although \ac{NR}
simulations of \ac{BNS} mergers have made tremendous progress in
recent years~\cite{Dietrich:2015pxa, Kastaun:2013mv, Dietrich:2017aum,
  Kastaun:2014fna, Tacik:2015tja, Paschalidis:2015mla, East:2015vix,
  Kastaun:2016elu, Kiuchi:2017pte, Radice:2013hxh, Bernuzzi:2012ci},
we do not yet have complete models of the inspiral, merger, and
post-merger regime, as we do for \ac{BBH}
systems~\cite{Taracchini:2013rva, Purrer:2015tud, Hannam:2013oca,
  Babak:2016tgq}.  The waveform models used for current \ac{GW}
analyses are either truncated prior to merger (this is the case for
all models used to analyse GW170817~\cite{Abbott:2018wiz}), or are
\ac{BBH} models through the merger and ringdown (which are included in
this study).  While one might expect that these approximations do not
impact parameter estimates, because the signal detectable by current
ground-based detectors contains negligible power at merger
frequencies, this assumption must be properly validated, especially in
light of the fact that the \ac{GW} energy emitted during the
post-merger stage can even exceed the \ac{GW} energy released during
the entire inspiral up to merger, cf.\ Fig.~3~\cite{Zappa:2017xba}.

For our study, we produce complete inspiral, merger and post-merger
\ac{BNS} waveforms by combining state-of-the-art tidal \ac{EOB}
waveforms for the inspiral, and \ac{NR} simulations of the late
inspiral and merger. We do this for two choices of the \ac{NS}
\ac{EOS}: a \emph{soft} \ac{EOS}, namely SLy~\cite{Douchin:2001sv},
corresponding to relatively compressible nuclear matter, and a
\emph{stiff} \ac{EOS}, namely MS1b~\cite{Mueller:1996pm},
corresponding to relatively incompressible nuclear matter.  These
yield \acp{NS} with low and high tidal deformabilities, respectively.
The two waveforms are then individually injected into a fiducial data
stream of the LIGO-Hanford and LIGO-Livingston
detectors~\cite{TheLIGOScientific:2014jea}.  We assume two different
noise \acp{PSD}, one from the first observing run of the LIGO
detectors\footnote{The \ac{PSD} is generated from 512\,s of LIGO data
  measured adjacent to the coalescence time of the first \ac{BBH}
  detection~\cite{Abbott:2016blz, abbott:2016wiq}. This is of
  comparable sensitivity to that of the LIGO detectors during both the
  first and second observing runs.} and another that is the projected
noise curve in the \ac{ZDHP} \cite{ZDHP} for the Advanced LIGO
detectors, although no actual noise is added to the data. The
LIGO-Virgo parameter-estimation algorithm {\ttfamily
  LALInference}~\cite{Veitch:2014wba,LSC}) is then employed to extract
the binary properties from the signal.  To determine the importance of
tidal effects in parameter-estimation analyses, we filter the data
with a variety of theoretical waveforms, with and without tidal
effects.  By measuring the \ac{SNR} of the post-merger part of the
signal, we determine the importance of the post-merger regime.

We describe the employed waveform models in Sec.~\ref{sec:BNSwfs} and
the construction of the hybrid waveforms in
Sec.~\ref{sec:injected-wf}. The parameter estimation methodology is
outlined in Sec.~\ref{sec:Bayes}. Our results are presented in
Sec.~\ref{sec:results}.

\section{Binary Neutron Star Waveforms}\label{sec:BNSwfs}

\subsection{Main features}

There are two main differences between \acp{GW} emitted from the
coalescence of \ac{BBH} and \ac{BNS} systems: (i) the presence of
tidal effects during the inspiral and (ii) a post-merger \ac{GW}
spectrum that might differ significantly from a simple
\ac{BH}-ringdown.

Considering the quasi-circular inspiral of two \acp{NS}, the emitted
\ac{GW} signal is chirp-like and characterized by an increasing
amplitude and frequency, similarly to the case of a \ac{BBH}
coalescence.  However, the deformation of the \acp{NS} in the external
gravitational field of the companion adds tidal information to the
\ac{GW}~\cite{Damour:1983b, Hinderer:2009ca, Damour:2009wj,
  Binnington:2009bb}.  Although tidal interactions (for non-spinning
binaries) enter the phase evolution at the 5PN
order~\cite{Damour:1990pi, Damour:1991yw, Damour:1992qi,
  Damour:1993zn, Racine:2004hs,Damour:2009wj, Hinderer:2009ca,
  Vines:2011ud}, the imprint on the \ac{GW} phase is visible even at
\ac{GW} frequencies $\lesssim150$\,Hz, e.g.~\cite{Hinderer:2009ca,
  Dietrich:2018uni}.  Closer to merger, tidal effects become stronger
and dominate the evolution~\cite{Bernuzzi:2014kca, Harry:2018hke}.

The magnitude of the tidal interaction is regulated by a set of tidal
deformability coefficients
\begin{equation}
\Lambda_\ell^{A,B} = \frac{2k_\ell^{A,B}}{C_{A,B}^{2\ell+1}(2\ell-1)!!},
\end{equation}
where ${A, B}$ label the two \acp{NS}, and $k_{\ell}^{A,B}$ and
$C_{A,B}$ denote their Love numbers and
compactnesses~\cite{Hinderer:2007mb, Damour:2009vw,Binnington:2009bb}.
Since the $\Lambda_\ell^{A,B}$'s depend on the internal structure of
the \acp{NS}, their measurement provides constraints on the \ac{EOS}
of cold degenerate matter at supranuclear densities.  For the two
equal-mass ($M_{A,B} = 1.35M_\odot$) configurations considered in this
article, the dominant, quadrupolar tidal deformabilities are
$\Lambda_2^{A,B} = 392.3$ and $\Lambda_2^{A,B} = 1536.4$ for the SLy
and the MS1b \ac{EOS}, respectively.  The tidal deformabilities of the
individual stars $\Lambda_2^{A,B}$ are difficult to measure, but the
combination
\begin{align}
\tilde\Lambda = \frac{16}{13} 
\frac{ (M_A+12M_B)M_A^4\Lambda_2^A}{ (M_A+M_B)^5 } + [A \leftrightarrow B] \,. \label{eq:tildeLambda}
\end{align}
can be extracted from the detected \ac{GW} signal with significantly
higher precision~\cite{Favata:2013rwa, Wade:2014vqa}.  $\tilde\Lambda$
captures the entire 5\ac{PN} tidal correction; it also enters at
6\ac{PN} order in linear combination with
\begin{align}
\delta\tilde\Lambda &= \left( M_A^2 - \frac{7996}{1319}M_AM_B - \frac{11005}{1319}M_B^2\right)
\frac{M_A^4\Lambda_2^A}{ (M_A+M_B)^6 } \nonumber\\
&- [A \leftrightarrow B] \,,
\label{eq:deltaTildeLambda}
\end{align}
which, however, is unlikely to measured by Advanced LIGO/Virgo
detectors~\cite{Wade:2014vqa}.

Extracting $\tilde{\Lambda}$ from a detected signal requires reliable
waveform models that accurately incorporate tidal effects.  Over the
last years, there have been improvements in the construction of
inspiral \ac{BNS} waveform approximants.  In \ac{PN}
theory~\cite{Blanchet:2013haa} several attempts have been made to
increase the known \ac{PN} order of tidal effects,
e.g.~\cite{Racine:2004hs, Hinderer:2009ca, Vines:2011ud}. Current
analytical knowledge includes (although incomplete) information up to
relative 2.5PN order~\cite{Damour:2012yf}.  While \ac{PN} based models
are computationally cheap, it has been shown that they are generally
unable to describe the binary coalescence in the late-inspiral, close
to the moment of merger~\cite{Bernuzzi:2012ci, Favata:2013rwa,
  Wade:2014vqa}.  Following the \ac{EOB}
approach~\cite{Buonanno:1998gg, Damour:2009wj} \ac{PN} knowledge can
be used in a resummed form to allow a more accurate description of the
binary evolution.  Indeed, the development of tidal \ac{EOB}
approaches has seen several improvements in recent years, showing
generally a good agreement with full \ac{NR} simulations up to the
moment of merger~\cite{Damour:2009wj, Baiotti:2010xh, Bernuzzi:2012ci,
  Bernuzzi:2014owa, Hotokezaka:2015xka, Hotokezaka:2016bzh,
  Hinderer:2016eia, Dietrich:2017feu}.  Very recently,
phenomenological prescriptions of tidal effects fitted to
\ac{PN}/\ac{EOB}/\ac{NR} have been proposed~\cite{Dietrich:2017aum,
  Kawaguchi:2018gvj, Dietrich:2018uni}.  These phenomenological tidal
descriptions can augment \ac{BBH} approximants to mimic \ac{BNS}
waveforms up to the moment of merger.

\ac{NR} simulations are necessary to describe the \ac{GW} signal
emitted after the merger of the two stars.  In general, the merger
remnant has a characteristic \ac{GW} spectrum with a small number of
broad peaks in the $f_\text{GW}\sim1.8-4$\,kHz frequency range. The
main peak frequencies of the post-merger \ac{GW} spectrum correlate
with properties of a zero-temperature spherical equilibrium
star~\cite{Bauswein:2011tp,Bauswein:2012ya} following
\ac{EOS}-independent quasi-universal relations~\cite{Bauswein:2011tp,
  Bauswein:2012ya, Hotokezaka:2013iia, Bauswein:2014qla,
  Takami:2014zpa, Bauswein:2015yca, Takami:2014tva, Bernuzzi:2015rla,
  Rezzolla:2016nxn, Bose:2017jvk}.  While measuring the post-merger
\ac{GW} signal would in principle allow one to determine the \ac{EOS}
independently of the inspiral signal, there is currently no waveform
approximant determining the phase evolution of the post-merger
waveform.  Independent of this, there have been approaches to obtain
information from the post-merger \ac{GW} signal, without using
waveform models, e.g.~\cite{Clark:2015zxa, Bose:2017jvk,
  Chatziioannou:2017ixj}.  Despite these advancements, there has been
no study to quantitatively establish whether the usage of a pure
inspiral \ac{GW} signal might lead to systematic biases or
uncertainties in determining the binary source properties.

\subsection{Waveform approximants}

The waveform approximants which we use in our study are 
described in the remainder of this section. 

\textbf{TaylorF2:} The \TaylorF{} model is a frequency-domain
\ac{PN}-based waveform model for the inspiral of \ac{BBH} systems.  It
uses a 3.5\ac{PN} accurate point-particle
baseline~\cite{Sathyaprakash:1991mt} and includes the spin-orbit
interaction up to 3.5\ac{PN}~\cite{Bohe:2013cla} and the spin-spin
interaction up to 3\ac{PN}~\cite{Arun:2008kb, Mikoczi:2005dn,
  Bohe:2015ana, Mishra:2016whh, Poisson:1997ha}.

\textbf{TaylorF2$_{\rm Tides}$:} The \TaylorFT{} uses \TaylorF{} as
baseline, but adds tidal effects up to 6PN as presented in
Ref.~\cite{Vines:2011ud}.  This model was used in the analysis of
GW170817~\cite{TheLIGOScientific:2017qsa, Abbott:2018wiz}.

\textbf{IMRPhenomD:} \IMRPhenomD{} is a phenomenological,
frequency-domain waveform model discussed in detail in
Refs.~\cite{Husa:2015iqa,Khan:2015jqa}. It describes non-precessing
\ac{BBH} coalescences throughout inspiral, merger, and ringdown.
While the inspiral is based on the \TaylorF{} approximation, it is
calibrated to \ac{EOB} results, and the late inspiral, merger and
ringdown are calibrated to \ac{NR} simulations.

\textbf{SEOBNRv4\_ROM:} This approximant is based on an \ac{EOB}
description of the general-relativistic two-body
problem~\cite{Buonanno:1998gg, Buonanno:2000ef}, with free
coefficients tuned to \ac{NR} waveforms~\cite{Buonanno:1998gg,
  Bohe:2016gbl}.  It provides inspiral-merger-ringdown waveforms for
\ac{BBH} coalescences.  For a faster computation of individual
waveforms we employ reduced-order-modeling techniques (indicated by
the suffix {\ttfamily ROM} in the name tag) ~\cite{Purrer:2015tud}.
  
\textbf{IMRPhenomD\_NRtidal:} To obtain \ac{BNS} waveforms, we augment
the \IMRPhenomD{} BBH approximant with tidal phase corrections.  The
NRtidal phase corrections have been introduced in
Ref.~\cite{Dietrich:2017aum} and combine \ac{PN}, \ac{EOB}, and
\ac{NR} information in a closed-form expression. The waveform model
terminates at the end of the inspiral; the termination frequency is
prescribed by fits to \ac{NR} simulations [see \cite{Dietrich:2018uni}
for details].  \IMRPhenomDNRtidal{} was also used in the LIGO-Virgo
analysis of GW170817~\cite{TheLIGOScientific:2017qsa, Abbott:2018wiz}.

\textbf{SEOBNRv4\_ROM\_NRtidal:} Similarly to the \IMRPhenomDNRtidal{}
model, this model augments the BBH approximant \SEOBNRROM{} with
NRtidal phase corrections~\cite{Dietrich:2017aum,Dietrich:2018uni}.

\textbf{TEOBResum:} The \TEOBResum{} model covers the low-frequency
regime when building the hybrid waveforms used in this study (see
Sec.\,\ref{Subsection:EOBdata}).  \TEOBResum{} was introduced in
\cite{Bernuzzi:2014owa} following the general formalism outlined in
\cite{Damour:2009wj}. The approximant incorporates an enhanced
attractive tidal potential derived from resummed \ac{PN} and
gravitational self-force expressions of the \ac{EOB} $A$-potential
that describes tidal interactions \cite{Damour:2009wj,Bini:2014zxa}.
The resummed tidal potential of \TEOBResum{} improves the description
of tidal interactions near the merger with respect to the
next-to-next-to-leading-order tidal \ac{EOB} model
\cite{Damour:2009vw,Bernuzzi:2012ci} and is compatible with in large 
regions of the \ac{BNS} parameter space with
high-resolution, multi-orbit \ac{NR} results within their
uncertainties \cite{Bernuzzi:2014owa,Dietrich:2017feu}.  The model as
employed in this article is restricted to irrotational \acp{BNS}.
  
\textbf{TEOBResum\_ROM:} Given the high computational cost of the
\TEOBResum{} approximant, we employ a reduced-order-model technique
when using this approximant in parameter
estimation~\cite{Lackey:2016krb}. There are some systematics
difference between \TEOBResum{} approximant and \TEOBResumROM{}, which
are discussed in the results section of Ref.~\cite{Lackey:2016krb}.

\section{Hybrid Waveforms}\label{sec:injected-wf}
\begin{figure*}[t]
  \includegraphics[width=0.98\textwidth]{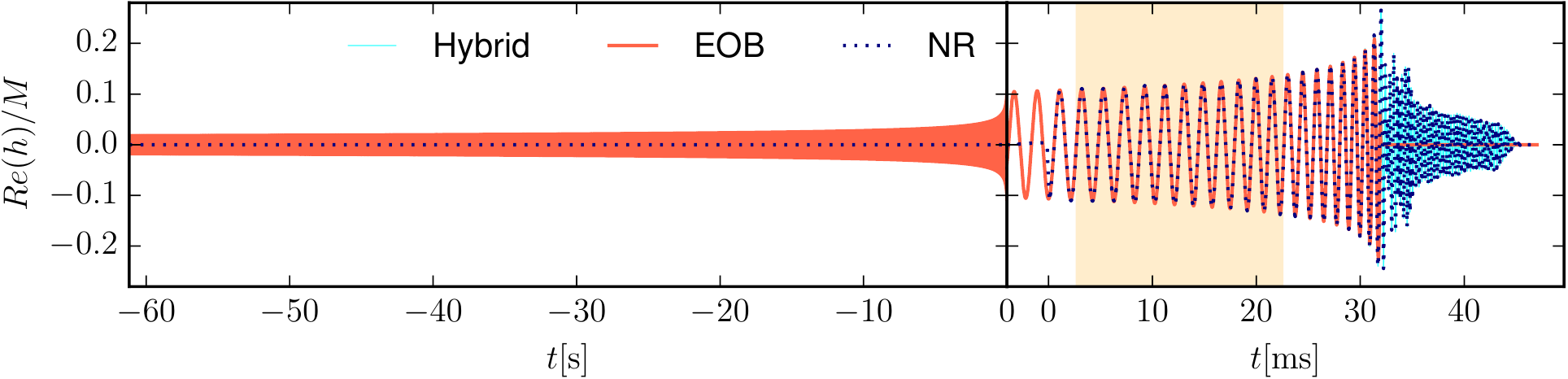}
  \caption{A hybrid waveform used in this study, with
    $M_A=M_B=1.35\,M_\odot$ and employing the SLy \ac{EOS}. The hybrid
    (thin, cyan line) consists of a tidal \ac{EOB} part (red) and an
    \ac{NR} part (dotted blue). The alignment interval is marked by
    the yellow shaded region in the right panel. The time $t=0$
    denotes the start of the \ac{NR} simulation. \label{fig:hybrid} }
\end{figure*}

In order to make meaningful statements for our study of tidal effects
and the post-merger waveform, it is necessary to construct a full
\ac{BNS} waveform that covers the inspiral, merger and post-merger
regimes.  We do this by following the procedure outlined in
Ref.~\cite{Dietrich:2018uni,Dietrich:2018phi}.  We combine analytical
waveforms constructed within the \ac{EOB} approach with waveforms
produced by \ac{NR} simulations.  The tidal \ac{EOB} part is computed
using the \TEOBResum{} model~\cite{Bernuzzi:2014owa} covering the
long, quasi-adiabatic inspiral portion of the signal (left panel of
Fig.~\ref{fig:hybrid}).  The \ac{NR} part covers the late inspiral,
merger, and post-merger regimes (right panel of
Fig.~\ref{fig:hybrid}).

Among the waveform models which we use for parameter estimation, the
ones that include tidal effects (\TaylorFT{}, \IMRPhenomDNRtidal{},
\SEOBNRROMNRtidal{}) differ from the \TEOBResum{} model, as they are
either purely \ac{PN}-based models, or phenomenological/\ac{EOB}-based
\ac{BBH} models to which tidal terms have been added following the
approach outlined in~\cite{Dietrich:2017aum,Dietrich:2018uni}. Using a
number of different waveform approximants allows us to estimate
systematic errors in the parameter estimation pipelines.  In addition,
we also use \TEOBResumROM{}~\cite{Lackey:2016krb} for parameter
estimation in order to check that the parameters of the injected
waveforms and their recovered values are consistent.  We refer the
reader also to a recent study of systematic effects presented in the
appendix of Ref.~\cite{Abbott:2018wiz} where pure tidal \ac{EOB}
waveforms using the \texttt{SEOBNRv4T}
model~\cite{Hinderer:2016eia,Steinhoff:2016rfi} were injected and
recovered.  We note that these injected waveforms lacked a post-merger
part and that the study restricted the characterization of systematic
effects to tidal waveform models, since no recovery with pure \ac{BBH}
approximants was carried out.  However, it has been generally found
that the \TaylorFT{} approximant predicts larger tidal deformabilities
than the \texttt{NRTidal} models, which we can verify within our
extended study.

In the following, we provide details about the construction of the
hybrid waveforms, including the discussion about the \TEOBResum{}
model and employed \ac{NR} data.

\subsection{EOB waveform}\label{Subsection:EOBdata}
We use the \TEOBResum{} waveform model in the frequency regime from
$30$\,Hz to $\sim$500$\,$Hz.  \TEOBResum{} is determined by seven
input parameters: the binary mass-ratio $q$ and the $\l=2,3,4$ tidal
polarizability parameters $\kappa^{A,B}_\ell$. The latter are related
to the $\Lambda^{A,B}_\ell$ tidal deformability parameters by
\begin{align}\label{kappa_barlam}
  \kappa^A_\ell &= q^{-1}\ X_A^{2\ell+1}\ (2\ell-1)!! \ \Lambda^A_\ell ,\\
  \kappa^B_\ell &= q \ X_B^{2\ell+1}\ (2\ell-1)!! \ \Lambda^B_\ell \ ,
\end{align}
with $X_{A,B}=M_{A,B}/(M_A+M_B)$.  For our equal-mass ($M_A=M_B=1.35$)
SLy and MS1b fiducial \acp{BNS}, one has $\kappa^A_2 = \kappa_2^B = 36.7749$ 
and $\kappa^A_2 = \kappa_2^B = 144.0378$ respectively.

Using the publicly available \TEOBResum
code\footnote{\url{https://bitbucket.org/account/user/eob_ihes/projects/EOB}},
we generate waveforms starting at a frequency of $30$\,Hz, which
corresponds to $\sim$60 seconds before the time of merger. We restrict
our analysis to the dominant $(2,2)$ mode throughout the paper.

\subsection{Numerical-relativity data}\label{Subsection:NRdata}

The numerical simulations were performed with the BAM code
\cite{Brugmann:2008zz, Thierfelder:2011yi}, which solves the Einstein
equations using the Z4c decomposition~\cite{Bernuzzi:2009ex,
  Hilditch:2012fp}, and have been previously published in
Ref.~\cite{Bernuzzi:2014owa}. The \ac{NR} data are publicly available
at \url{http://www.computational-relativity.org/},
cf.~\cite{Dietrich:2018phi}.

The two binary configurations used in this work describe equal-mass
\ac{BNS} systems with a total mass of $2.70\,M_\odot$, {\it i.e.}, a
chirp mass $\mathcal{M} =
(M_AM_B)^{3/5}/(M_A+M_B)^{1/5}=1.1752\,M_\odot$. The two waveforms
differ in their choice of the \ac{EOS} modeling the supranuclear
matter inside the \ac{NS}.  The \ac{NR} waveforms start at an initial
dimensionless frequency $(M_A+M_B)\omega_{22} = 0.038$, which
corresponds to $455$\,Hz. They cover $\sim$10 orbits prior to merger,
the merger itself, and post-merger. The merger frequencies $f_{\rm
  merger}$ are $2010$\,Hz and $1405$\,Hz for waveforms with the SLy
and the MS1b \ac{EOS}, respectively, and the frequency content of the
post-merger signal reaches up to $\sim 4000$\,Hz. At the moment of
merger, the phase uncertainty as estimated in~\cite{Bernuzzi:2014owa}
is $\Delta \phi = \pm 0.40\,$rad for the SLy and $\Delta \phi = \pm
3.01\,$rad for the MS1b setup.  The larger phase uncertainty of the
MS1b setup gets partially compensated for by the fact that this setup
has also significantly larger tidal effects due to the stiffer
\ac{EOS}.  For a more detailed discussion about uncertainties in
\ac{NR} simulations, we refer to Ref.~\cite{Bernuzzi:2016pie}.

\subsection{Hybrid waveform}\label{Subsection:Hybrid}

To hybridize the tidal \ac{EOB} and \ac{NR} waveforms modeling the
same physical \ac{BNS} system, we first align the two waveforms. This
is done by minimizing
\begin{equation}
  \mathcal{I}(\delta t,\delta \phi) = \int_{t_i}^{t_f}dt|\phi_{\rm NR}(t) - \phi_{\rm EOB}(t+\delta t) + \delta \phi|^2 ,
\end{equation}
with $\delta \phi$ and $\delta t$ being relative phase and time
shifts.  $\phi_{\rm NR}$ and $\phi_{\rm EOB}$ denote the phases of the
\ac{NR} and tidal \ac{EOB} waveform, respectively.  The alignment is
done in a time window $[t_i,t_f]$ that corresponds to the
dimensionless frequency window $[0.04, 0.06]$.  Previous comparisons
have shown that in this interval the agreement between the \ac{NR} and
\ac{EOB} waveforms is excellent~\cite{Bernuzzi:2014owa,
  Hotokezaka:2015xka, Dietrich:2017feu}.  Additionally, our particular
choice for this window allows us to average out the phase oscillations
linked to the residual eccentricity $(∼10^{-2})$ of the \ac{NR}
simulations.

Once the waveforms are aligned, they are stitched together by the
smooth transition
\begin{align}
& h_{\rm Hyb}(t) \nonumber \\ \quad & = \left\{
\begin{small}
  \begin{array}{ll}
    h_{\rm EOB}(t')e^{i\phi} & : t \le t_i\\
    h_{\rm NR}(t)H(t) + h_{\rm EOB}(t')e^{i\phi}[1- H(t)] & : t_i \le t \le t_f\\
    h_{\rm NR}(t) & : t \ge t_f
  \end{array}
\end{small}
\right.
\end{align}

where $t' = t + \delta t$, and $H(t)$ is the Hann window function
\begin{equation}
  H(t) := \frac{1}{2} \bigg[ 1 - \cos \bigg( \pi \frac{t-t_i}{t_f - t_i} \bigg) \bigg]\,.
\end{equation}

\begin{figure}[!t]
  \centering  \includegraphics[width=0.5 \textwidth]{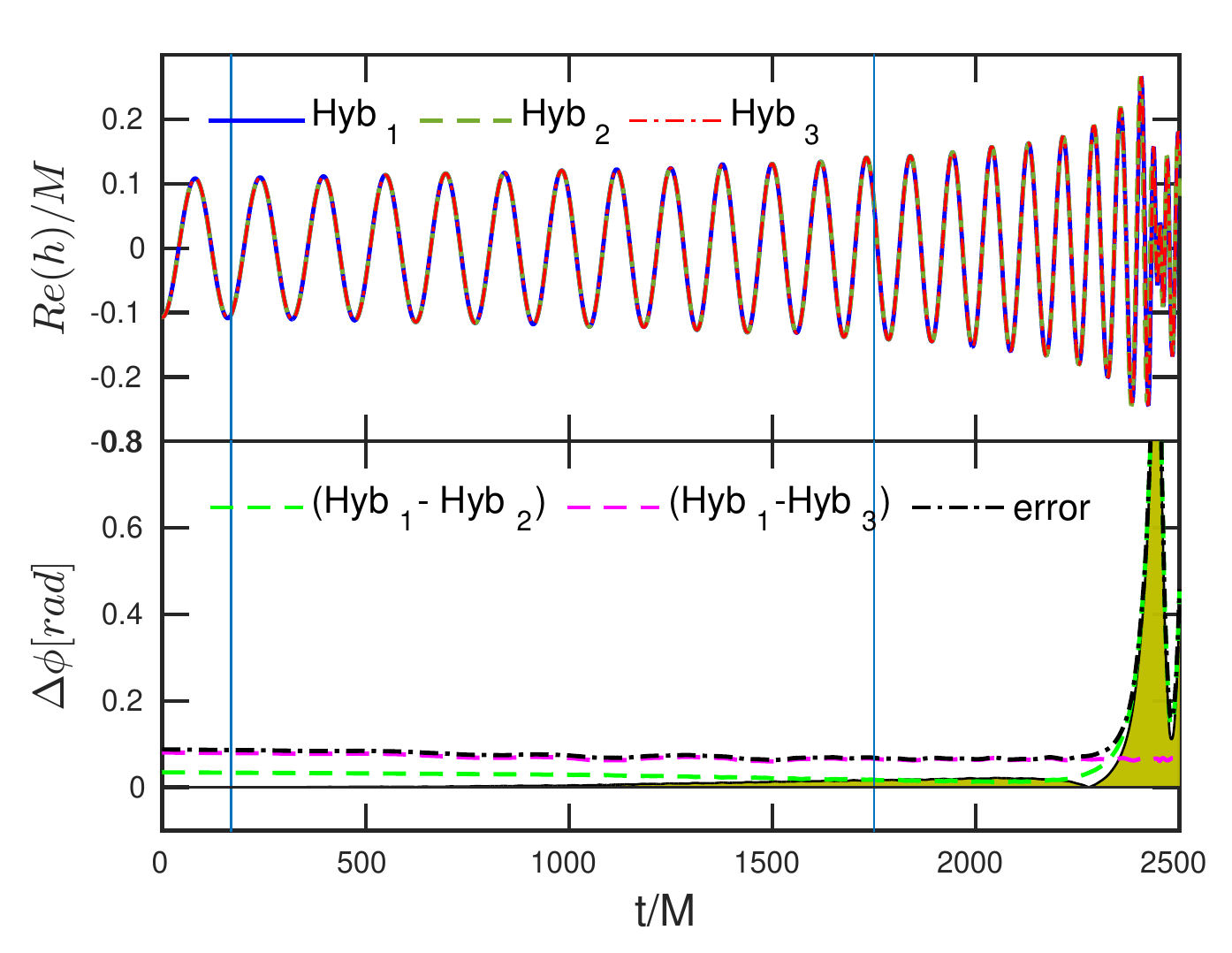}
  \caption{\label{fig:hybrid_accu} (\emph{Top panel}) The three
    flavors of the hybrid discussed in Sec.\,\ref{Subsection:Hybrid},
    each constructed with a different \ac{NR} resolution and \ac{EOB}
    integrator settings. Red and blue dashed curves represent the same \ac{NR}
    resolution but a different \ac{EOB} integrator setting. The green dashed
    curve has a lower \ac{NR} resolution but same \ac{EOB} integrator setting
    as for the red curve. The hybrid in Fig.\,\ref{fig:hybrid}
    corresponds to the blue curve here. (\emph{Bottom panel}) Phase
    difference between the hybrid of Fig.\,\ref{fig:hybrid} and its
    two other realizations. The green dashed curve represents the
    absolute dephasing with the hybrid with low \ac{NR} resolution,
    while the pink curve shows the absolute dephasing with the hybrid
    using a different \ac{EOB} integrator setting. The black curve represents
    the absolute error defined in Eq.\,(\ref{eqn:err}). The vertical
    lines mark the boundaries of the alignment window. The olive shaded
    region is the dephasing between two NR resolutions used.}
\end{figure}

To estimate the uncertainty of the hybrid waveform, we present in
Fig.~\ref{fig:hybrid_accu} the same configuration, but evolved with
different resolutions for the \ac{NR} simulation and different time
resolutions $\rm dt$ for the \ac{ODE} integrator used in the
\TEOBResum model.  We denote the high resolution \ac{NR} simulations
in which $\sim 128$ grid points cover the \ac{NS} and in which we use
$\rm dt=0.50$ for the \ac{EOB} as $\rm Hyb_1$ (blue line in the top
panel).  $\rm Hyb_2$ is the hybrid employing a lower resolution
\ac{NR} dataset with $\sim 96$ grid points covering the \ac{NS}, but
the same resolution for the \ac{EOB} \ac{ODE} integrator. $\rm Hyb_3$
is the hybrid with \ac{NR} resolution of 128 grid points and $\rm dt=
0.25$ for the \ac{EOB} \ac{ODE} integrator resolution. To check the
accuracy of the hybrid, we compute the dephasing between these three
cases and an error is shown in the bottom panel of
Fig.\,\ref{fig:hybrid_accu}. We define the error as
\begin{equation}
{\rm err} = \sqrt[]{(\phi_{{\rm{Hyb_1}}} -\phi_{{\rm{Hyb_2}}})^2+(\phi_{{\rm {Hyb_1}}} -\phi_{{\rm{Hyb_3}}})^2}\,.
\label{eqn:err}
\end{equation}
Here, $\phi_{\rm{Hyb}_1}$, $\phi_{\rm{Hyb}_2}$ and $\phi_{\rm{Hyb}_3}$
are the phases of hybrids $\rm Hyb_1$, $\rm Hyb_2$, and $\rm Hyb_3$,
respectively. For the error computation, we aligned all three hybrids
within the frequency interval [32, 34]Hz and then calculate the phase
differences.  We find that the difference between $\rm Hyb_1$ and $\rm
Hyb_2$ is below $ \sim 0.1$ radian and at merger well within the
\ac{NR} uncertainty (olive shaded region).  The effect of the \ac{ODE}
integration within the \ac{EOB} model is even smaller. However, this
study does not include the systematic effects of the underlying
\ac{EOB} model, see Ref.~\cite{Nagar:2018zoe} for further details.

\section{Bayesian inference}\label{sec:Bayes}
In this section, we provide a brief overview of the Bayesian inference
setup we use to determine the physical properties of the injected
signal.  The time series of detector data, $d(t)$, can be modeled as
the sum of the true \ac{GW} signal and detector noise, denoted by
$h(t)$ and $n(t)$, respectively:
\begin{equation}
  d(t) = h(t) + n(t).
\end{equation}
Under this assumption, we can use Bayes' theorem to determine the
posterior probability density $p(\boldsymbol{\theta} | d(t))$ of the
parameters $\boldsymbol{\theta}$, given the data $d(t)$ as
\begin{equation}
  p(\boldsymbol{\theta} | d(t)) \propto \mathcal{L}(d(t)| \boldsymbol\theta) p(\boldsymbol{\theta}) \,,
  \label{eqn:bayes}
\end{equation}
where $\mathcal{L}(d(t)| \boldsymbol\theta)$ is the likelihood, or the
probability of observing the data $d(t)$ given the signal model
described by $\boldsymbol\theta$, and $p(\boldsymbol{\theta})$ denotes
the prior probability density of observing such a source.  For
Gaussian noise, the likelihood for a single detector is given
by~\cite{Finn:1992wt}
\begin{equation}
  \mathcal{L}(d(t)|\boldsymbol{\theta}) \propto \exp\left[\ -2 \int_0^\infty \frac{|\tilde{d}(f) - \tilde{h}(f,\boldsymbol{\theta})|^2}{S_{\rm det}(f)} df\right]\ ,
\end{equation}
where tildes denote Fourier transforms of time series introduced so
far and $S_{\rm det}(f)$ is the one sided \ac{PSD} of the detector.
Under the assumption that noise in different detectors is not
correlated, this expression is readily generalized to the case of a
coherent network of detectors by taking the product of the likelihoods
in each detector~\cite{Finn:1995ah}.

Credible intervals for a specific subset of source parameters in the
set $\boldsymbol{\theta}$ may be obtained by marginalising the full
posterior over all but those parameters.  Obtaining credible intervals
therefore requires sampling the multidimensional space of source
parameters.  We do this with {\ttfamily lalinference\_mcmc}, a
Markov-Chain Monte Carlo sampler algorithm~\cite{vanderSluys:2008qx}
included in the {\ttfamily LALInference} package~\cite{Veitch:2014wba}
as part of the LSC Algorithm Library (LAL)~\cite{LSC}.  In addition to
the chirp mass, $\mathcal{M}$, the binary mass ratio, $q=M_B/M_A(\leq
1)$, the dimensionless spin magnitudes of the two \acp{NS},
$\chi_{A,B}$, the tidal deformability parameters, $\tilde{\Lambda}$
and $\delta \tilde{\Lambda}$
[Eqs.\,(\ref{eq:tildeLambda})-(\ref{eq:deltaTildeLambda})], our
parameter space also includes the luminosity distance, an arbitrary
reference phase and time for the \ac{GW} signal, the inclination angle
of the binary with respect to the line-of-sight of the detectors, the
polarization angle, and the right ascension and declination of the
source, {\it i.e.}, its sky location.

A key ingredient of Bayesian analyses is one's choice of the prior
probability density $p(\boldsymbol{\theta})$ (along with the
hypothesis used to perform inference on the data, which in our
specific case is the choice of the waveform model).  We use a uniform
prior distribution in the interval $[1 M_\odot, 3 M_\odot]$ for the
component masses, and a uniform prior between $-1$ and $1$ for both
dimensionless aligned spins.  We also pick a uniform prior
distribution for the individual tidal deformabilities
$\Lambda_2^{A,B}$ between $0$ and $5000$ for all waveform models
except \TEOBResumROM, in which case they are bound between $50$ and
$5000$. With regards to tidal deformability, our setup makes the
simplifying assumption of ignoring correlations between $\Lambda_2^A$,
$\Lambda_2^B$, and the mass parameters, which are known to
exist~\cite{Flanagan:2007ix, Hinderer:2009ca, Favata:2013rwa}.  For
all other parameters we follow the setup of standard \ac{GW} analyses,
{\it e.g.,} \cite{Abbott:2018wiz}.

\section{Results}\label{sec:results}
\begin{table}
  \begin{tabular} {@{\hspace{.2cm}}l@{\hspace{.2cm}}c@{\hspace{.4cm}}c@{\hspace{.4cm}}c@{\hspace{.4cm}}c@{\hspace{.4cm}}c@{\hspace{.4cm}}c@{\hspace{.2cm}}}
    \toprule[1.pt]
    \toprule[1.pt]
    \addlinespace[0.3em]
    EOS & $\tilde{\Lambda}$ & $\mathcal{M}$ $[M_\odot]$  & $\chi_{\rm eff}$ & PSD & $f_{\rm high}$ & SNR \\[0.2em]
    \addlinespace[0.2em]
    \midrule[0.75pt]
    \addlinespace[0.2em]
    \multirow{4}{*}{SLy} & \multirow{4}{*}{392} & \multirow{4}{*}{1.1752} & \multirow{4}{*}{0.0} & \multirow{2}{*}{O1} & 2048\,Hz & 25\\
    & & & & & 2048\,Hz & 100 \\
    \cmidrule[0.3pt]{5-7}
    & & & & \multirow{2}{*}{ZDHP} & 2048\,Hz &  100\\
    & & & & & 8192\,Hz & 100\\
    \addlinespace[0.2em]
    \cmidrule[0.3pt]{1-7}
    \addlinespace[0.2em]
    \multirow{4}{*}{MS1b} & \multirow{4}{*}{1536} & \multirow{4}{*}{1.1752} & \multirow{4}{*}{0.0} & \multirow{2}{*}{O1} & 2048\,Hz & 25\\
    & & & & & 2048\,Hz & 100 \\
    \cmidrule[0.3pt]{5-7}
    & & & & \multirow{2}{*}{ZDHP} & 2048\,Hz &  100\\
    & & & & & 8192\,Hz & 100\\
    \bottomrule[1.pt]
    \bottomrule[1.pt]
  \end{tabular}
  \caption{The eight injections used in this study.  We consider two equal-mass, non-spinning \ac{BNS} hybrid waveforms (see Sec.\,\ref{sec:injected-wf}) with the same chirp mass, but different \ac{EOS}, and hence different tidal polarizability $\tilde{\Lambda}$.  The \ac{SNR} of each injection is specified in the seventh column.  When assessing the impact of tidal effects on the analysis of a \ac{GW} inspiral signal, we use the first observing run's noise \ac{PSD}.  In such cases, the waveform models employed to recover the signal are \IMRPhenomDNRtidal, \SEOBNRROMNRtidal, \TaylorFT, \IMRPhenomD, \TaylorF, and \TEOBResumROM.  When assessing the impact of post-merger dynamics on \ac{GW} inference, instead, we use the projected noise curve for the Advanced LIGO detectors in the \ac{ZDHP} configuration, and \IMRPhenomDNRtidal and \TaylorFT waveform models for the signal recovery.  Given that the high-frequency content of the post-merger portion of the signal reaches $\sim 4$\,kHz, we produce injections with sampling rates of $16384$\,Hz and $4096$\,Hz, and correspondingly use a high-frequency cutoff in our Bayesian analysis of $f_{\rm high}=8192$\,Hz and $f_{\rm high}=2048$\,Hz. The merger frequencies are $2010$\,Hz and $1405$\,Hz for waveforms with SLy and MS1b \ac{EOS}, respectively.}
  \label{tab:injections}
\end{table}

We consider a two-detector network that consists of the LIGO
interferometers situated in Hanford and Livingston, US.  We inject the
equal-mass SLy and MS1b hybrid \ac{GW} signals discussed in
Sec.\,\ref{sec:injected-wf}~\cite{Schmidt:2017btt}.  For each hybrid
signal, we consider $4$ different injection setups, for a total of $8$
scenarios, as summarized in Table \ref{tab:injections}.  To study the
impact of neglecting/including tidal effects in the analysis of a
\ac{BNS} \ac{GW} inspiral signal, we assume a noise \ac{PSD} from the
first observing run of the LIGO detectors and perform two injections
per hybrid, with \acp{SNR} $25$ and $100$ [rows 1, 2, 5, and 6 in
Table \ref{tab:injections}].  The former value allows us to address a
scenario in which the source is detected with a moderately high
\ac{SNR}, namely $\sim 3$ times the detection threshold.  An \ac{SNR}
of $100$ is chosen, instead, to assess the impact that tidal effects
have on the recovery of \ac{BNS} source properties for an extreme
scenario.  The same, high value of \ac{SNR} is used when assessing the
impact of post-merger dynamics on \ac{GW} inference.  For this scope,
we assume the projected noise curve for the Advanced LIGO detectors in
the \ac{ZDHP} \cite{ZDHP}.  Because the high-frequency content of the
post-merger portion of the \ac{GW} reaches $\sim 4000$\,Hz, we produce
injections with two different sampling rates: $16384$\,Hz and
$4096$\,Hz, which correspond to a high-frequency cutoff in our
Bayesian analysis of $f_{\rm high}=8192$\,Hz and $f_{\rm
  high}=2048$\,Hz, respectively (rows 3, 4, 7, and 8 in Table
\ref{tab:injections}).  In all cases, no actual noise is added to the
data.  This allows us to obtain posteriors that do not depend upon a
specific noise realization, therefore isolating systematic errors.
All injections start at a frequency of $30$\,Hz, while our Bayesian
analysis uses a low cutoff frequency $f_{\rm low} = 32$\,Hz to
generate template waveforms.  As discussed in Sec.\,\ref{sec:BNSwfs},
we use a number of different waveform approximants for estimating
parameters; this allows us to qualitatively assess systematic errors
present in such waveform models.  As opposed the full
inspiral-merger-post--merger \ac{BNS} signals we inject, the waveform
models used for estimating parameters are limited to the inspiral
regime.

The insights we gained by comparing the results of parameter
estimation using different waveform models are summarized in the
following subsections.

\subsection{Effects of tidal terms}\label{subsection:results-tidal}

We first assess under what conditions neglecting tidal effects incurs
a bias in the recovered masses and spins, and then investigate
configurations where uncertainties in the modeling of tidal effects
can incur a bias in both the masses and spins, and the measurement of
the tidal deformability.

\begin{figure}[!t]
    \includegraphics[width=0.50\textwidth]{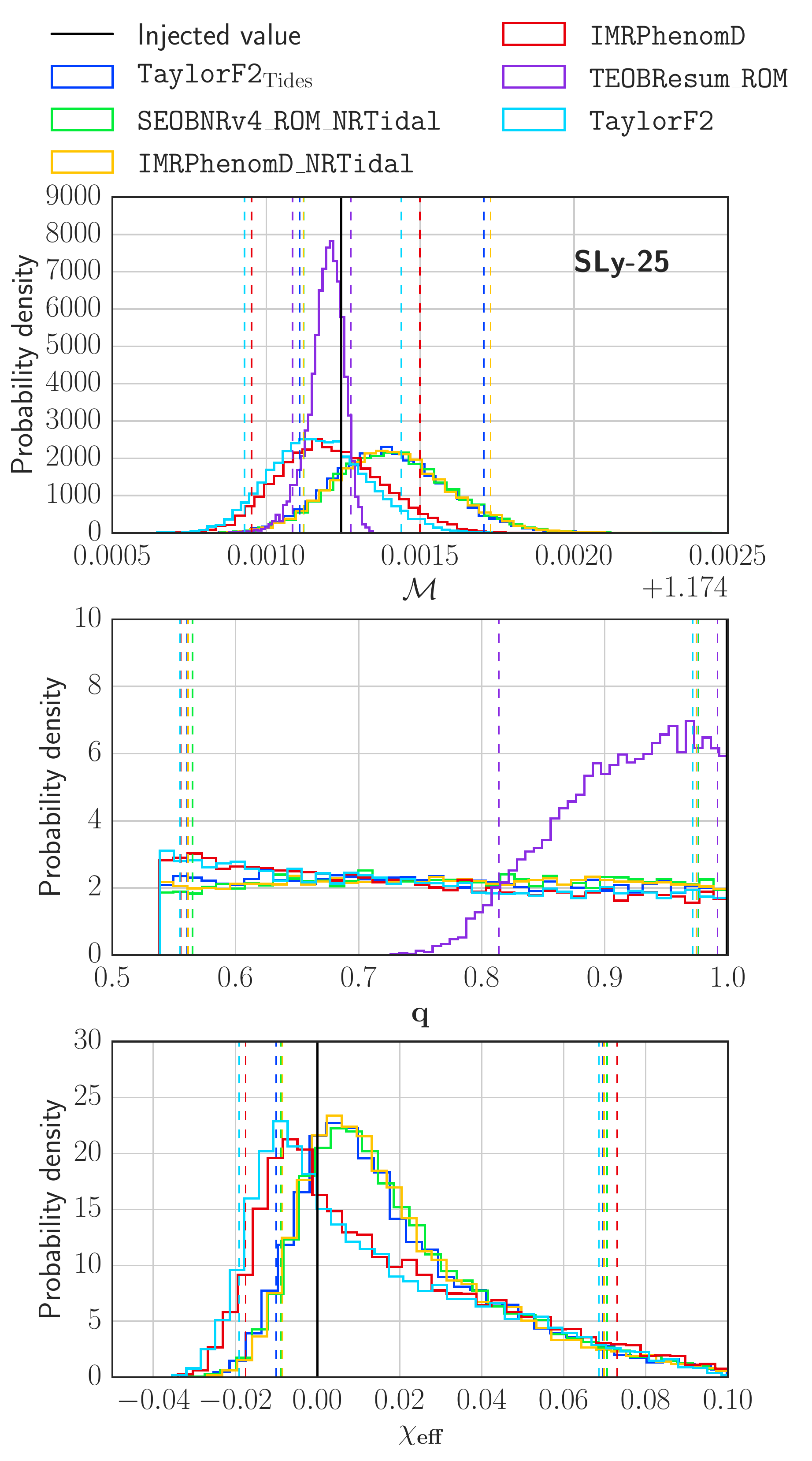}
    \caption{The chirp mass $\mathcal{M}$ ({\it top panel}), mass
      ratio $q$ ({\it middle panel}), and effective spin $\chi_{eff}$
      ({\it bottom panel}) posterior distributions for the \ac{BNS}
      injection with SLy \ac{EOS} at \ac{SNR} 25. The vertical dashed
      lines mark the $90$\% credible intervals, while the solid black
      line indicates the injected value. }
  \label{fig:SLy25}
\end{figure}

The waveform phase evolution is dominated by the chirp mass
\cite{Peters:1963ux}, followed by the mass-ratio $q$, and then spin
effects. The spins are characterized in parameter measurements by a
weighted sum of the two spins, $\chi_{\rm eff}$ \cite{Ajith:2009bn},
\begin{equation}
  \chi_{\rm eff} = \frac{M_A \chi_A + M_B \chi_B}{M_A+M_B},
\end{equation}
which is related to the leading-order \ac{PN} spin contribution to the
waveform phase \cite{Kidder:1992fr}. There is a partial signal
degeneracy between the mass-ratio and the spin \cite{Cutler:1994ys,
  Poisson:1995ef, Baird:2012cu, Hannam:2013uu, Ohme:2013nsa}, in that
increased spin can be compensated by a lower mass ratio; the
degeneracy is not exact because mass-ratio and spin effects enter at
different \ac{PN} order. Tidal effects enter at yet higher (5\ac{PN})
order, but can also be partially mimicked by a change in mass ratio
and spin; thus, neglecting tidal effects could lead to a bias in
$\chi_{\rm eff}$ and in the mass ratio. We will investigate the extent
of these biases in the following, by considering the recovery of the
chirp mass, $\mathcal{M}$, the mass-ratio $q$, and $\chi_{\rm eff}$.

We first consider the SLy configuration observed with an \ac{SNR} of
25; note that the GW170817 observation was at a comparable \ac{SNR} of
32.4, and the SLy configuration has a tidal deformability of
$\tilde{\Lambda} = 392$, also consistent with the measurements of
GW170817~\cite{Abbott:2018wiz}.  Figure \ref{fig:SLy25} shows the
chirp mass $\mathcal{M}$, mass-ratio $q$ and effective spin $\chi_{\rm
  eff}$.  We see that the injected value of the chirp mass is inside
the 90\% credible interval for all approximants used to recover the
injected signal.  The chirp-mass (and mass-ratio) posterior
distribution yielded by the \TEOBResumROM approximant is significantly
different from the other ones.  This is due to the fact that this
waveform model is non-spinning; the parameter estimation algorithm
therefore explores a parameter space that differs from the one it
handles in the case of all other approximants.  With the exception of
\TEOBResumROM, the peak of the chirp-mass distribution is slightly
higher than the injected chirp-mass value for approximants that
include tidal terms, but it is shifted to lower values for the
approximants that neglect tidal terms.  The degeneracy between
mass-ratio and spin leads to a relatively flat distribution in $q$ for
all approximants; although there is a hint that the posterior is
starting to rail against the lower prior limit of $q$ in the parameter
estimation run, the effect is small.  The effect of this degeneracy is
most clearly illustrated by the results from the \TEOBResumROM
approximant, which, once more, is a non-spinning approximant, and
without this degree of freedom the mass ratio is recovered with far
greater accuracy.  In measurements of the spin, the peak of the
distribution is again slightly shifted away from the injected value,
towards lower values for non-tidal approximants and towards higher
values for tidal approximants.  Once again, the injected values lie
within the 90\% credible interval.

\begin{figure}[!t]
  \includegraphics[width=0.50\textwidth]{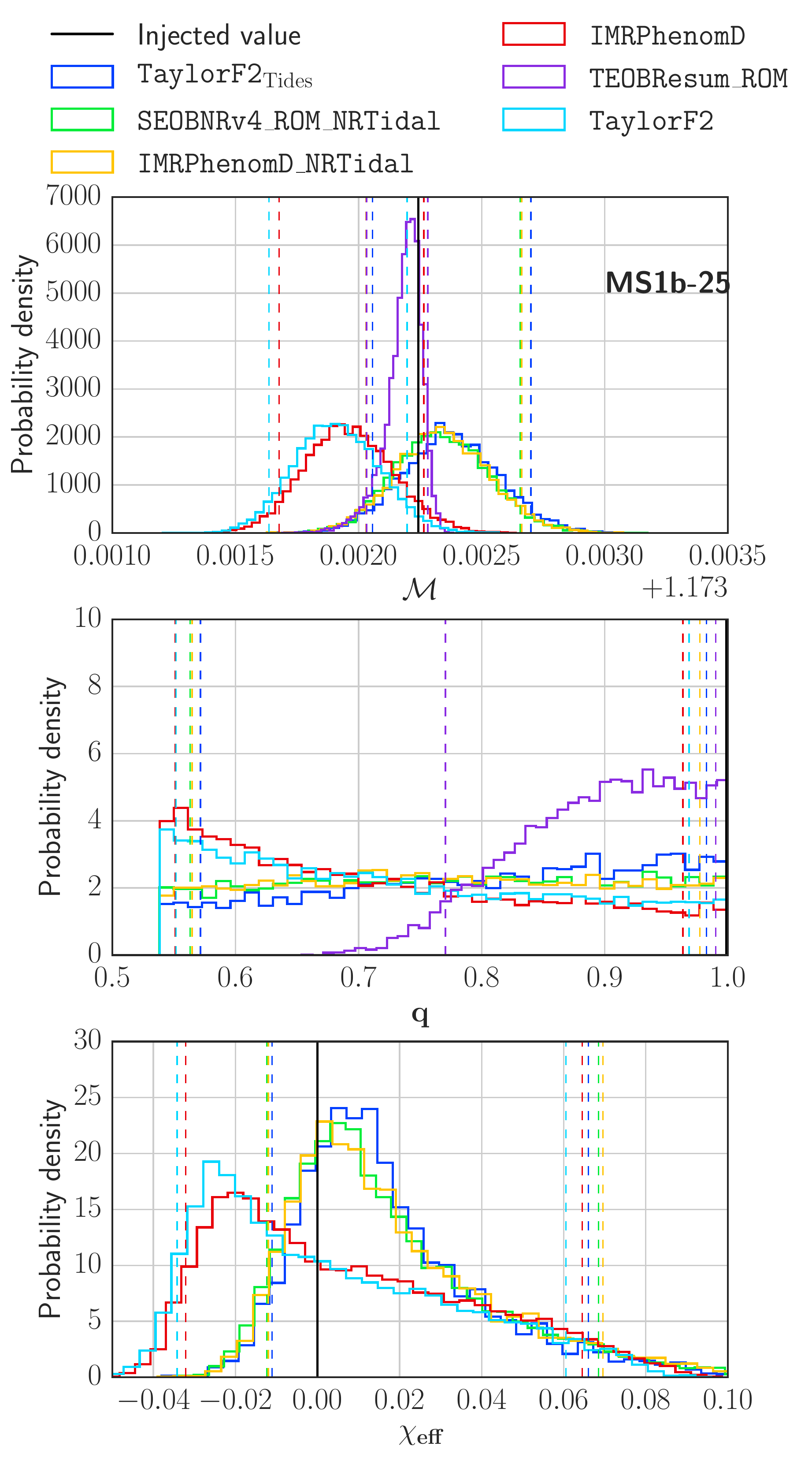}
  \caption{Same as Fig.\,\ref{fig:SLy25} but assuming the MS1b
    \ac{EOS} to produce the injected \ac{BNS} signal. }
  \label{fig:MS1b25}
\end{figure}

\begin{figure}[!ht]
  \includegraphics[width=0.50\textwidth]{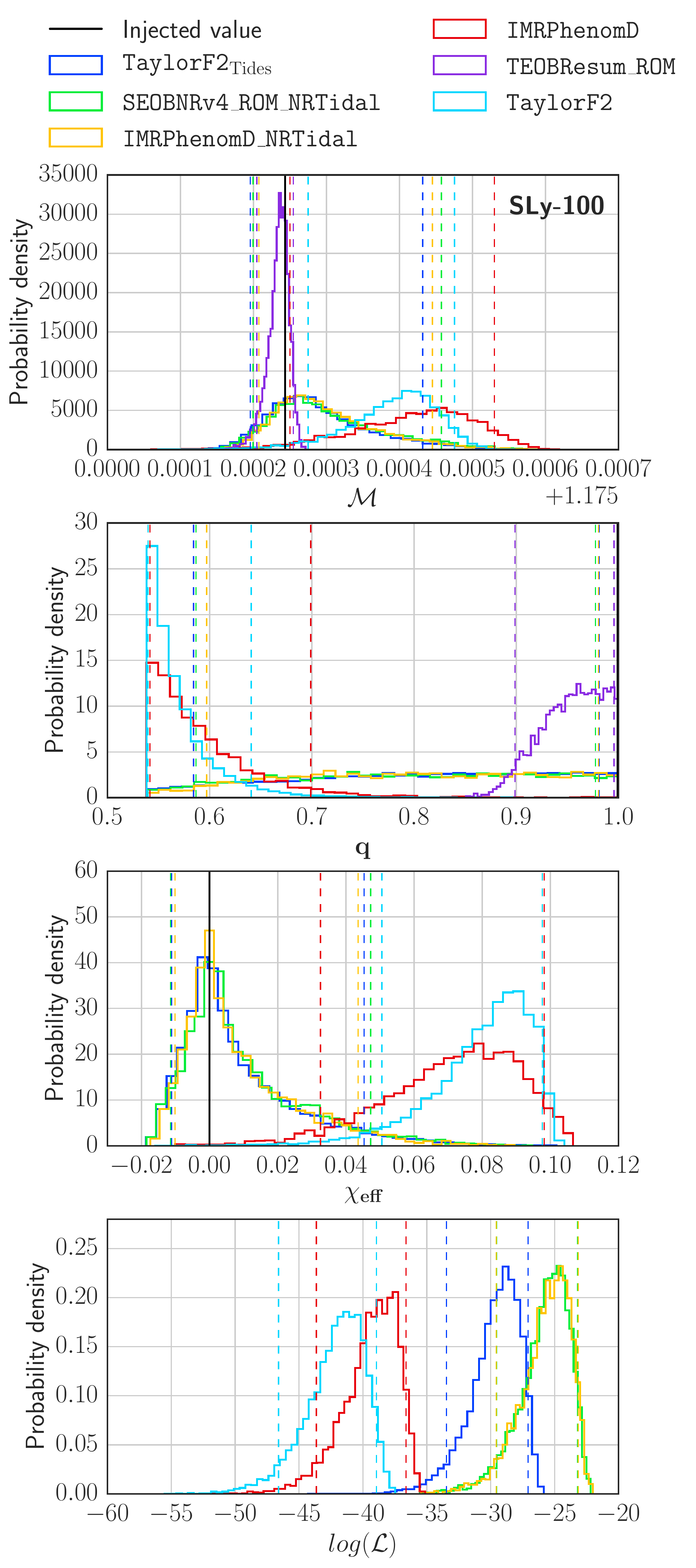}
  \caption{Same as Fig.\,\ref{fig:SLy25}, but with an injected
    \ac{SNR} of 100.  We also show the distribution of the logarithm
    of the likelihood in the bottom panel.}
  \label{fig:SLy100}
\end{figure}

We conclude that for soft \acp{EOS} at this \ac{SNR}, neglecting tidal
terms does not lead to a significant bias in measurements of masses
and spins.

This picture changes when we consider a stiffer \ac{EOS}. Figure
\ref{fig:MS1b25} shows the same quantities, but for the MS1b
configuration ($\tilde\Lambda=1536$) injected at \ac{SNR} 25. Although
the 90\% credible intervals for the mass ratio and spin agree with the
injected values, the bias in the measurement of the chirp mass
obtained with non-tidal approximants is more significant: the injected
value is outside the 90\% credible interval for \TaylorF, and it is
very close to the edge of the 90\% credible interval for
\IMRPhenomD. Further, the peak of the mass-ratio distribution is
railing more significantly against the lower prior limit in the case
of the non-tidal approximants, and, similarly, we observe an increase
in the shift of the peak of the spin distribution for the same
waveform models.  Collectively, these results indicate that for stiff
\acp{EOS} neglecting tidal terms \emph{does} lead to a bias in the
measurement of the masses and spins.

Assuming future observations will be similar to GW170817
($\tilde{\Lambda}<720$~\cite{Abbott:2018wiz}) and that \acp{SNR}
higher than 25 will be rare, our results suggest that neglecting tidal
effects will not significantly bias measurements of masses and spins
for typical observations.  However, careful analyses will be required
once individual events are combined to extract information about the
\ac{BNS} population as a whole.

\begin{figure*}[!t]
  \includegraphics[width=1.0\textwidth]{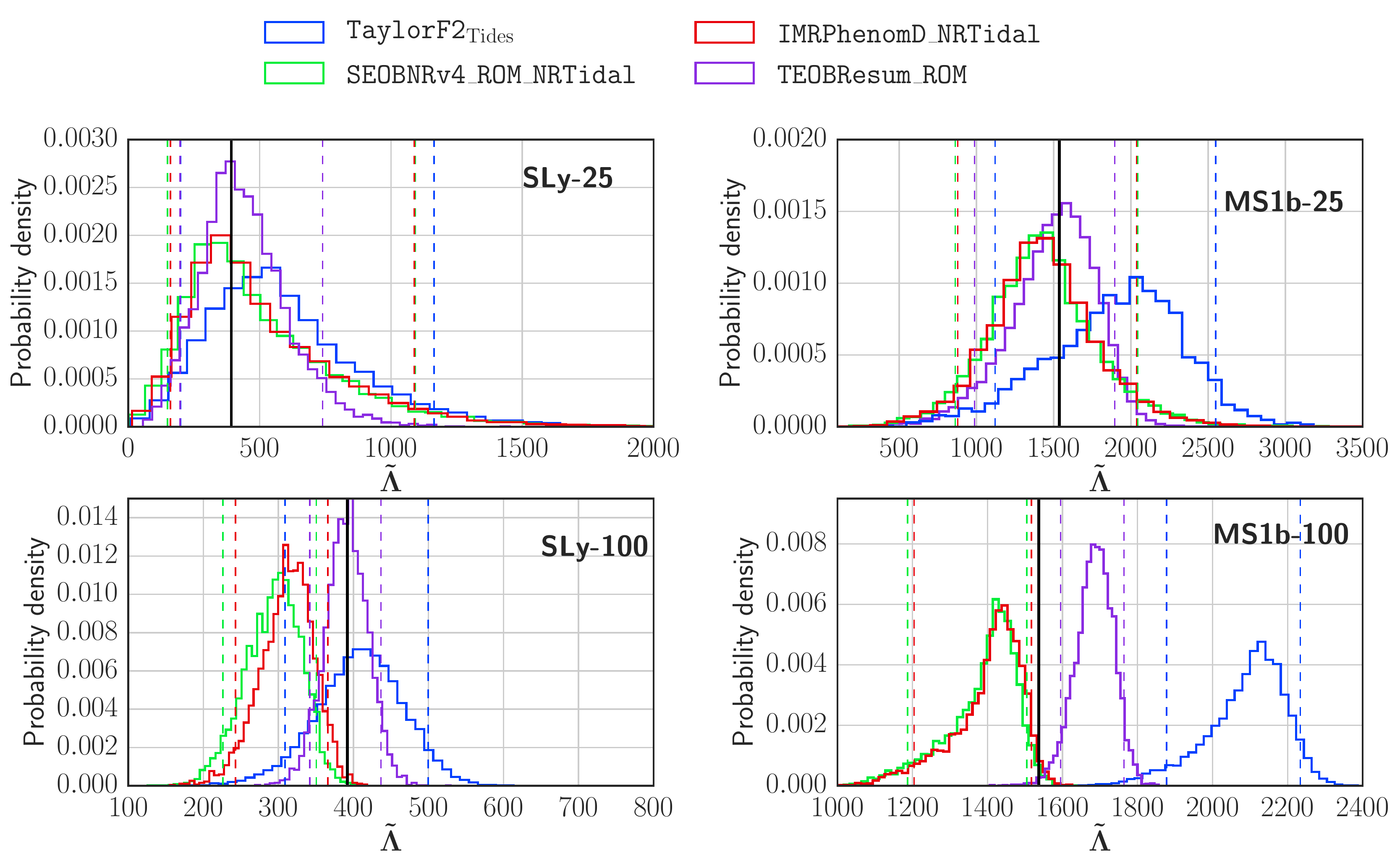}
  \caption{Measurements of the tidal deformability parameter. The two
    panels on the left show results for the SLy signal injected at
    \ac{SNR} 25 ({\it top}) and 100 ({\it bottom}).  The two panels on
    the right show results for the MS1b signal injected at \ac{SNR} 25
    ({\it top}) and 100 ({\it bottom}).}
  \label{fig:lambda}
\end{figure*}

We also investigate how these results change for a much higher
\ac{SNR}. Figure \ref{fig:SLy100} shows measurements of the chirp
mass, mass ratio and spin for the SLy configuration, now injected at
an \ac{SNR} of 100. We see that the chirp mass is now biased away from
the correct value for the two approximants that do not include tidal
terms. When estimating the parameters, we enforced a limit that
$m_{1,2} \in [1,3]M_\odot$, which implies $\mathcal{M}_c \in [0.8706,
2.6117]M_\odot$.  The parameter estimation code adjusts the masses and
spins to find the best match with the data, and for those approximants
that do not include tidal terms, the search rails against the limits
on the masses, as well as on the physical limit $\chi \leq 1$ for the
spins. This is most clear in the plot of the posterior distribution
for $q$. The Figure also includes a plot of the logarithm of the
likelihood; we see that the \TaylorF and \IMRPhenomD approximants
cannot be made to match the data as accurately as the tidal
approximants, and so their likelihoods are lower. (More generous
limits on the masses in the parameter recovery may lead to a higher
likelihood for these approximants, but we do not expect it to be as
high as for the approximants that include tidal terms, since biases in
the masses and spins can only partially mimic the missing tidal
effects.) We also see that the \TaylorFT approximant, although it
contains tidal terms, is not as accurate as the \NRtidal approximants,
for which the tidal terms have been tuned to \ac{NR} waveforms.

We now move on to measurements of the tidal deformability,
$\tilde{\Lambda}$, for which we can compare the accuracy of different
tidal approximants.  In Fig.\,\ref{fig:lambda}, the left two panels
show the results for SLy injections (at \acp{SNR} 25 and 100), and the
two right panels show the results for MS1b.  The results shown here
are entirely consistent with the systematics tests performed for the
LIGO-Virgo Collaboration analysis of the properties of
GW170817~\cite{Abbott:2018wiz}.  In particular, all tidal approximants
agree within their 90\% credible intervals at \acp{SNR} measured to
date, for both soft and stiff \acp{EOS}, and for all configurations
the \TaylorFT approximant can be used to place an upper bound on
$\tilde{\Lambda}$, as in Refs.~\cite{TheLIGOScientific:2017qsa,
  Abbott:2018wiz}.

It is interesting to note, however, the at an \ac{SNR} of 100, the
measurement using the \NRtidal approximants is biased away from the
injected value of the hybrid, which was constructed from the
\TEOBResum approximant. This is an indication that we do not have
sufficient control over systematics for high-\ac{SNR} setups. A
possible explanation for this behavior is that tidal effects in the
\NRtidal model are larger than in the \TEOBResum model used to produce
the injected signals, as already highlighted in Fig.\,10 of
Ref.\,\cite{Dietrich:2018uni}. Another possible explanation is that
this is due to differences between the \NRtidal and \TEOBResum
approximants in the \ac{BH} limit. The agreement of \TaylorFT in the
bottom-left panel of Fig.\,\ref{fig:lambda} is accidental.  (We
believe that this is due to a compensation of two effects: \TaylorFT
models underestimating tidal effects and therefore overestimating
$\tilde{\Lambda}$, and systematics errors in the point-particle
description~\cite{DudiInPrep}.) We note also that at \ac{SNR} 100, the
\TEOBResum approximant provides a biased measurement for MS1b. This
may be surprising at first, since \TEOBResum was used in the
construction of the MS1b hybrid, but the approximant is used only up
to the hybridisation frequency, from which point onwards the \ac{NR}
waveform is used. There is an \ac{SNR} of $\sim 16$ from the
hybridization frequency up to the merger. As already stated, there are
also systematic differences between \TEOBResum and
\TEOBResumROM~\cite{Lackey:2016krb}.  We found a phase difference
between our hybrid and the waveform generated using \TEOBResumROM of
about $\sim 4$\,rad at merger for the MS1b case and $\sim 0.8$\,rad
for the SLy case. We suggest this to be the reason for the offset
between the injected value and the \TEOBResumROM result at
${\rm\ac{SNR}}=100$, cf.~Fig.~\ref{fig:lambda}.

\subsection{Effect of post\-merger}\label{subsection:results-post-merger}
\begin{figure}[!t]
  \centering
  \includegraphics[width=0.50 \textwidth]{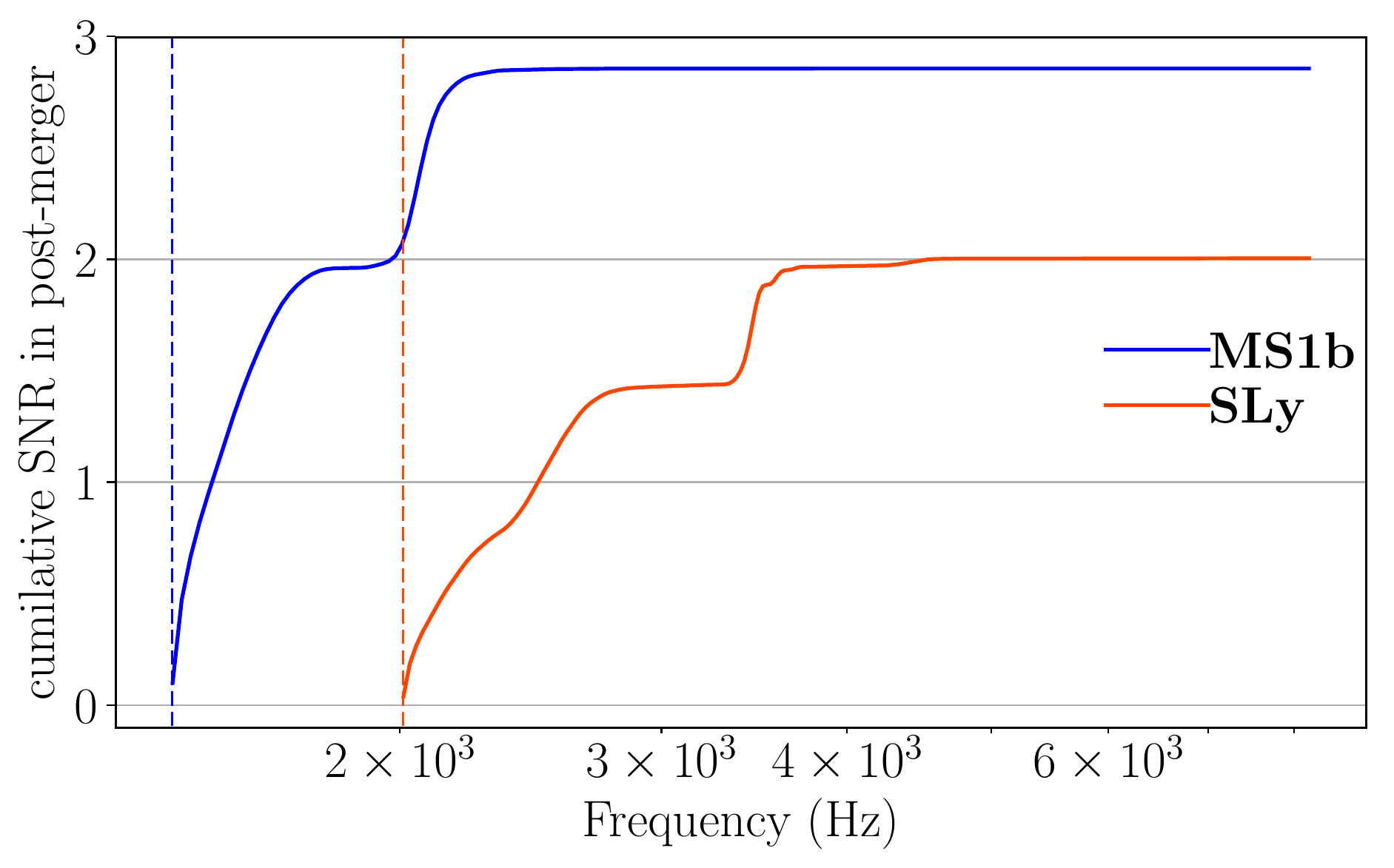}
  \caption{\ac{SNR} accumulated during the post-merger regime as a
    function of frequency. The vertical lines indicate the merger
    frequency, where the computation of the \ac{SNR} starts. Blue and
    red curves correspond to the hybrid being injected. The
    \ac{ZDHP} projected noise curve for the Advanced LIGO detectors is
    used.}
  \label{fig:pm-snr}
\end{figure}

We now investigate whether the lack of the post-merger part of the
signal in the models, which we use for Bayesian inference, could lead
to biases in parameter measurements. Previous studies of post-merger
\ac{GW} signals~\cite{Clark:2015zxa, Bose:2017jvk,
  Chatziioannou:2017ixj} suggest that this portion of the signal would
be detectable, and its properties measurable, only if the \ac{SNR} of
the post-merger regime alone were above $\sim 5$. Recently,
\citet{Chatziioannou:2017ixj} found that for soft \acp{EOS} an
\ac{SNR} of $3$--$4$ might be sufficient for a detection of the
post-merger \ac{GW} using the BayesWave
algorithm~\cite{Cornish:2014kda}.  In this article, we assume a
threshold \ac{SNR} of $5$ to produce more conservative estimates.
Figure~\ref{fig:pm-snr} shows the accumulated \ac{SNR} of the
post-merger regime for our SLy and MS1b configurations at a total
signal \ac{SNR} of 100. In order to achieve a post-merger \ac{SNR} of
5, we would need a total signal \acp{SNR} of approximately 185 and
250, for the MS1b and SLy \acp{EOS}, respectively.  These would
correspond to source distances of 17\,Mpc and 13\,Mpc.  If we assume
an \ac{SNR} signal detection threshold of 10 and a uniform volume
distribution of sources throughout the universe, then only about 1 in
every 6000 observations will have an \ac{SNR} greater than 185. This
suggests that it is unlikely for Advanced LIGO and Virgo detectors to
be able to measure post-merger signals, and that it is therefore
extremely unlikely that the post-merger part that is absent from our
signal models will bias the parameter recovery from the inspiral
waveform.  Nonetheless, our hybrid waveforms provide the opportunity
to conclusively test this expectation, and that is what we do in this
section.

\begin{figure}[!t]
  \centering
  \includegraphics[width=0.50 \textwidth]{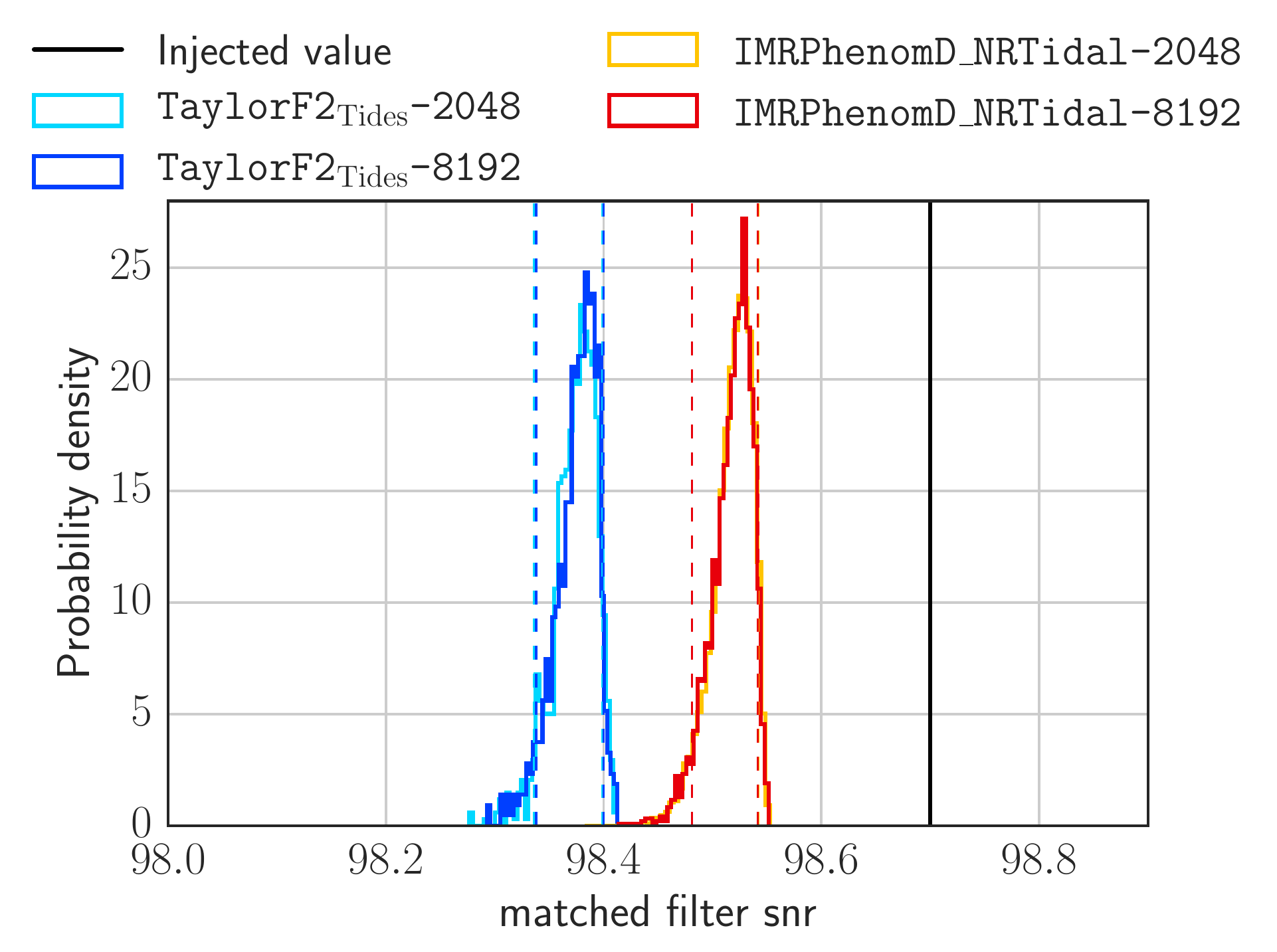}
  \caption{The matched filter \ac{SNR} recovered by \TaylorFT and
    \IMRPhenomDNRtidal for the \ac{EOS}-MS1b hybrid injected at
    \ac{SNR}=100, using two different sampling rates for the
    templates, namely, $4096$\,Hz and $16384$\,Hz.  Vertical dashed
    lines indicate $90\%$ credible intervals, while the black solid
    line marks the injected value.}
  \label{fig:optimal-snr}
\end{figure}

To quantify the impact of the post-merger portion of the signal, we
injected full, hybrid waveforms at ${\rm\ac{SNR}}=100$ and compared
results obtained using upper cutoff frequencies $f_{\rm high} =
2048$\,Hz and $f_{\rm high} = 8192$\,Hz. The merger frequencies
$f_{\rm merger}$ are $2010$\,Hz and $1405$\,Hz for waveforms with the
SLy and the MS1b \ac{EOS}, respectively, and the frequency content of
the post-merger signal reaches up to $\sim 4000$\,Hz, with peak
frequencies at $f_1$ $\sim 2600$\,Hz and $f_2$ $\sim 3400$\,Hz for
SLy, and $f_1$ $\sim 1600$\,Hz and $f_2$ $\sim 2100$\,Hz for MS1b.  We
find that the results of parameter estimation for the two different
cutoff frequencies are remarkably in agreement with each other,
suggesting that the lack of post-merger content in the models used for
parameter estimation has no impact on the recovery for this
configuration.  Figure \ref{fig:optimal-snr} shows that the recovered
\ac{SNR} is insensitive to the upper cutoff frequency, {\it i.e.}, to
the presence/absence of post-merger content in the injected signal,
and to the waveform model used in the recovery, since results for
\IMRPhenomDNRtidal and \TaylorFT are quite close.  Further, in all
four cases the full injected \ac{SNR} is recovered.  The $\sim 2$\%
drop from the nominal injected \ac{SNR} of 100 to ${\rm\ac{SNR}}=98$
is due to the fact that while the injected signal starts at $30$\,Hz,
the filtering against the template waveform has a minimum frequency of
$32$\,Hz.

\begin{figure}[h]
  \centering
  \includegraphics[width=0.5 \textwidth]{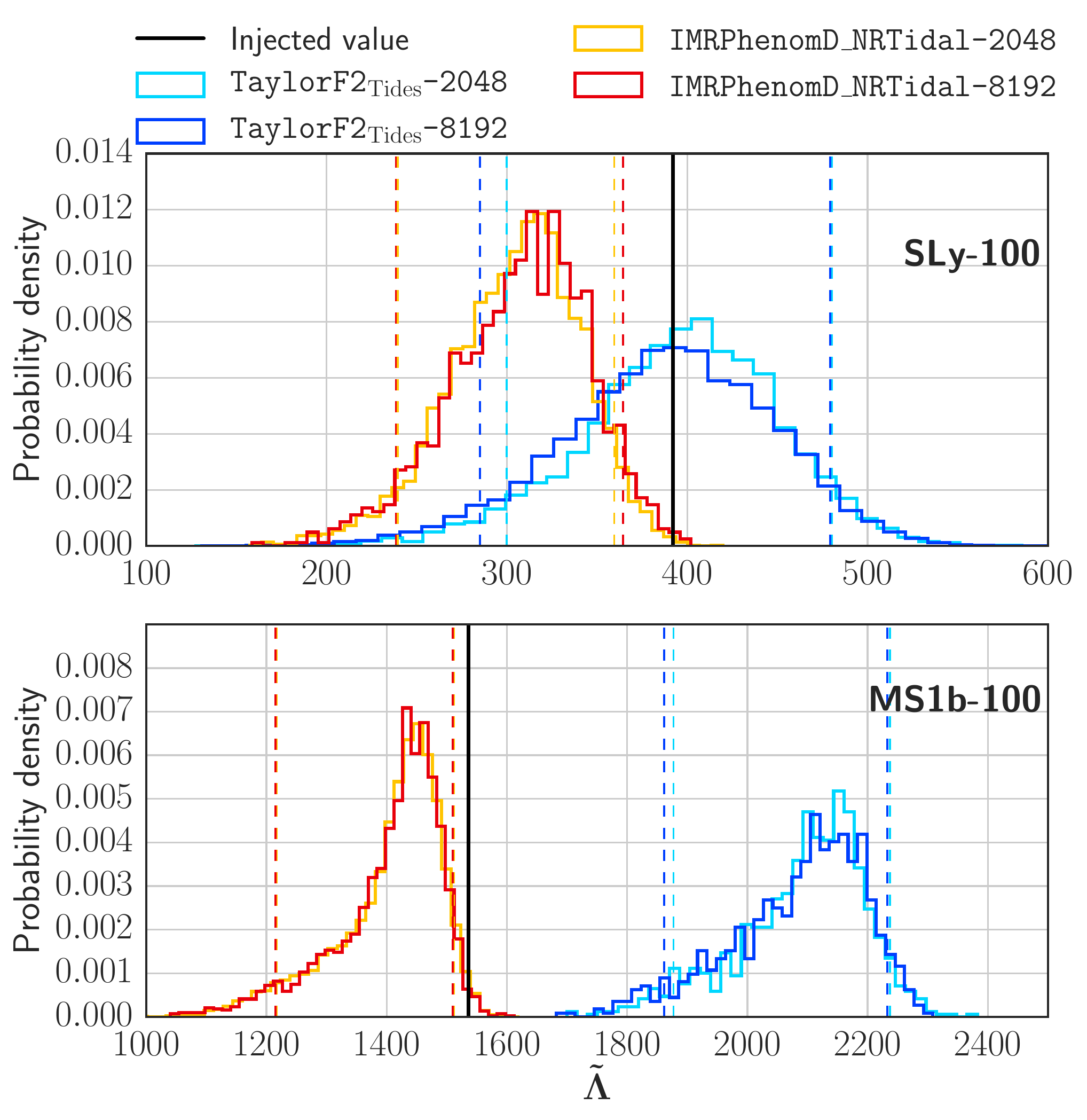}
  \caption{Tidal deformability posterior distributions found when
    injecting the SLy \ac{EOS} ({\it top panel}) and the MS1b \ac{EOS}
    ({\it bottom panel}) \ac{BNS} hybrids at \ac{SNR} $100$.  The
    recovery is performed the \IMRPhenomDNRtidal and \TaylorFT
    approximants with sampling rates $16384$\,Hz and $4096$\,Hz.
    Vertical dashed lines indicate $90\%$ credible intervals, while
    the black solid lines mark the injected values.}
  \label{fig:sly-post}
\end{figure}

Figure \ref{fig:sly-post} shows the posterior distribution for
$\tilde{\Lambda}$ for both choices of upper cutoff frequency, for the
SLy (upper panel) and MS1b (bottom panel) configuration injected at
$\rm\ac{SNR}=100$, using the \IMRPhenomDNRtidal and \TaylorFT waveform
models.  The upper cutoff frequency, or equivalently the inclusion or
absence of the post-merger regime in the injected signal, has
negligible effect on the results.  As was the case for the
${\rm{\ac{SNR}}}=100$ injections performed with the second observing
run noise \ac{PSD}, the \NRtidal models underestimate
$\tilde{\Lambda}$ for all injections. For an explanation, we refer the
reader to \ref{subsection:results-tidal}.

These results are consistent with our expectation that the post-merger
will have negligible impact on our parameter recovery with the current
generation of interferometric \ac{GW} detectors.

\section{Conclusion}
In this work, we considered two possible sources of systematic
uncertainties in the Bayesian parameter estimation of the \ac{GW}
signal emitted by a coalescing \ac{BNS}.  We focused on two questions:
(i) {\it What is the impact of neglecting tidal effects in the
  analysis of the inspiral \ac{GW} signal?} (ii) {\it Does the use of
  inspiral-only waveforms lead to a significant loss of information,
  or possibly to biases in the estimation of the source properties?}
To answer these questions, we produced complete \ac{BNS} \ac{GW}
signals by combining state-of-the-art \ac{EOB} waveforms for the
inspiral, and \ac{NR} simulations of the late inspiral and merger (see
Sec.~\ref{sec:injected-wf}), we injected such signals into fiducial
data streams of the LIGO detectors (see Table \ref{tab:injections}),
and, finally, we used the parameter-estimation algorithm {\ttfamily
  LALInference}~\cite{Veitch:2014wba, LSC}) to extract the source
properties from the data streams containing the injected signals.  We
addressed the first question by filtering the data with a variety of
theoretical waveforms, with and without tidal effects, whereas to
address the second question we quantified the importance of the
post-merger part of the signal by measuring its \ac{SNR}.

We showed that neglecting tidal effects in the inspiral waveforms used
to infer the source properties does not bias measurements of masses
and spin for a canonical observation at \ac{SNR}=25, as long as the
\ac{NS} \ac{EOS} is fairly soft ($\tilde{\Lambda}\lesssim400$).  In
the high \ac{SNR} regime and/or for stiff \acp{EOS}
($\tilde{\Lambda}\sim1500$), however, there will be a significant bias
in the measurements of masses and spins when inspiral waveform models
that do not include tidal effects are used (Figs.\,\ref{fig:MS1b25}
and \ref{fig:SLy100}).

We also found that the recovery of chirp mass $\mathcal{M}$,
mass-ratio $q$, effective spin $\chi_{\rm eff}$, and tidal
deformability $\tilde{\Lambda}$ is consistent among all tidal models
that include tidal effects for an injected \ac{SNR} of $25$ for both
soft and stiff \acp{EOS}.  In this context, \TaylorFT overestimates
the recovered value of $\tilde{\Lambda}$, as stated in the analysis of
GW170817~\cite{Abbott:2018wiz}. This is due to \TaylorFT favouring
larger values of $\tilde{\Lambda}$ in order to compensate for the
smaller tidal effects it includes in the phasing of the late inspiral
regime when compared to \NRtidal models~\cite{Dietrich:2018uni}. At
high \ac{SNR}, the impact of systematics present in the various
waveform models increases. In particular, the bottom panels of
Fig.\,\ref{fig:lambda} show that \NRtidal models yield a conservative
lower estimate of $\tilde{\Lambda}$. It is not clear whether the
differences between the \NRtidal and \TEOBResum approximants are
dominated by differences in the \ac{BH} ($\tilde{\Lambda} \rightarrow
0$) limit, or in the description of tidal effects, and this requires
further study.

Considering the possibility for upcoming detections with large
\acp{SNR} due to the increasing sensitivity of advanced \ac{GW}
detectors our study showed the importance to further improve \ac{BNS}
waveform models in coming years.

To investigate the impact of neglecting the post-merger portion of the
signal in waveform models used for parameter estimation of \ac{BNS}
signals, we calculated the \ac{SNR} of the post-merger regime.  As
suggested in Refs.~\cite{Clark:2015zxa, Bose:2017jvk,
  Chatziioannou:2017ixj} this part of the signal would be detectable,
and its properties measurable, only if its \ac{SNR} were above $\sim
5$.  As shown in Fig.\,\ref{fig:pm-snr}, achieving \ac{SNR} $\sim 5$
in the post-merger regime requires a total \ac{SNR} of about $\sim
200$.  The odds of having an event with \ac{SNR} $\sim 200$ are 1 in
every 6000 observations. This makes it unlikely for second generation
detectors to measure the post-merger part of \ac{BNS} \ac{GW} signals,
and thus the absence of the post-merger regime in waveform models
currently used in Bayesian inference is not worrisome for Advanced
LIGO and Virgo.

\begin{acknowledgments}
  It is a pleasure to thank Vivien Raymond and Lionel London for
  helping in setting up the NR infrastructure and parameter estimation
  runs. We also thank Yoshinta Setyawati and Sebastian Khan for useful
  discussions.
  R.D. and B.B. were supported in part by DFG grants GK 1523/2 and
  BR 2176/5-1.
  T.D. acknowledges support by the European Unions Horizon 2020
  research and innovation program under grant agreement No 749145,
  BNSmergers.
  M.H. and F.P. were supported by Science and Technology Facilities
  Council (STFC) grant ST/L000962/1 and European Research Council
  Consolidator Grant 647839.
  S.B. acknowledges support by the EU H2020 under ERC Starting Grant,
  no.~BinGraSp-714626.
  F.O. acknowledges supported by the Max Planck Society.
  NR simulations have been performed on the supercomputer SuperMUC
  at the LRZ (Munich).
  We are grateful for computational resources provided by Cardiff
  University, and funded by an STFC grant supporting UK Involvement in
  the Operation of Advanced LIGO.
  %
\end{acknowledgments}
\bibliography{refs.bib}
\end{document}